%% file: Manuscript.tex
\newif\ifdoubleblind
\newif\ifacm
	
\ifacm
	\documentclass[sigconf]{acmart}
\else
	\documentclass[conference]{IEEEtran}
\fi

\input{tex/pkg/packages}

\begin{document}

\input{tex/variables}
\input{tex/acro}

\input{tex/title}

\input{tex/abstract}
\input{tex/arxiv}

\maketitle

\input{tex/introduction}
\input{tex/relatedWork}

\input{tex/approach}
\input{tex/methodology}
\input{tex/results}

\input{tex/conclusion}

\input{tex/acknowledgment}

\ifacm
\else
	\bibliographystyle{IEEEtran}
	\bibliography{Bibliography}
\fi

\end{document}

%% file: tex/pkg/packages.tex
\usepackage{amsmath}

\usepackage{multirow}
\usepackage{tabularx,booktabs}
\usepackage{xspace}

\usepackage[binary-units=true]{siunitx}
\usepackage{tikz}
\usetikzlibrary{positioning}

\ifacm
\else
	\input{tex/pkg/packages_ieee}
\fi

%% file: tex/pkg/packages_acm.tex
\usepackage[nolist]{acronym}
\usepackage{booktabs} 
\usepackage{subfig}
\usepackage{balance}

\usepackage{array}
\usepackage{graphicx}
\usepackage{wasysym}
\usepackage{xparse}

\usepackage{makecell}
\newcolumntype{Y}{>{\centering\arraybackslash}X}

\hypersetup{draft}

\settopmatter{printacmref=false} 
\renewcommand\footnotetextcopyrightpermission[1]{} 
\pagestyle{plain} 


%% file: tex/pkg/packages_ieee.tex
\usepackage{graphicx}
\usepackage{epstopdf}
\usepackage{standalone}
\usepackage[nolist ]{acronym}
\usepackage{tcolorbox}
\usepackage{pdfpages}
\usepackage[bookmarks=false]{hyperref}
\usepackage{pdfcomment}
\usepackage{hologo}

\usepackage{xstring}

\usepackage[utf8]{inputenc}
\usepackage{marvosym}

\usepackage{algorithm}
\usepackage{algpseudocode}

\usepackage{psfrag}
\usepackage{graphicx}

\usepackage{rotating}

\renewcommand{\baselinestretch}{1}

\usepackage[caption=false,font=footnotesize]{subfig}
\usepackage{stfloats}

%% file: tex/variables.tex
\newcommand{\paperTitle}{A Low Cost Modular Radio Tomography System for Bicycle and Vehicle Detection and Classification}
\newcommand{\paperAuthors}{Marcus Haferkamp, Benjamin Sliwa and Christian Wietfeld}
\newcommand{\paperEmails}{$\{$Marcus.Haferkamp, Benjamin.Sliwa, Christian.Wietfeld$\}$@tu-dortmund.de}

\newcommand\single{1\textwidth}
\newcommand\double{.48\textwidth}
\newcommand\triple{.32\textwidth}
\newcommand\quarter{.24\textwidth}
\newcommand\singleC{1\columnwidth}
\newcommand\doubleC{.475\columnwidth}

\newcommand{\figurePadding}{0pt}
\newcommand{\figureTopPadding}{\figurePadding}
\newcommand{\figureBottomPadding}{\figurePadding}
\newcommand\red[1]{\colorbox{red}{#1}}

\newcommand\tikzFig[2]
{
	\begin{tikzpicture}
		\node[draw,minimum height=#2,minimum width=\columnwidth,text width=\columnwidth,pos=0.5]{\LARGE #1};
	\end{tikzpicture}
}

\newcommand{\dummy}[3]
{
	\begin{figure}[b!]  
		\begin{tikzpicture}
		\node[draw,minimum height=6cm,minimum width=\columnwidth,text width=\columnwidth,pos=0.5]{\LARGE #1};
		\end{tikzpicture}
		\caption{#2}
		\label{#3}
	\end{figure}
}

\newcommand\pos{h!tb}

\newcommand{\basicFig}[7]
{
	\begin{figure}[#1]  	
		\vspace{#6}
		\centering		  
		\includegraphics[width=#7\columnwidth]{#2}
		\caption{#3}
		\label{#4}
		\vspace{#5}	
	\end{figure}
}
\newcommand{\fig}[4]{\basicFig{#1}{#2}{#3}{#4}{0cm}{0cm}{1}}

\newcommand\sFig[2]{\begin{subfigure}{#2}\includegraphics[width=\textwidth]{#1}\caption{}\end{subfigure}}
\newcommand\vs{\vspace{-0.3cm}}
\newcommand\vsF{\vspace{-0.4cm}}

\newcommand{\subfig}[3]
{%
	\subfloat[#3]%
	{%
		\includegraphics[width=#2\textwidth]{#1}%
	}%
	\hfill%
}

\newcommand\circled[1] 
{
	\tikz[baseline=(char.base)]
	{
		\node[shape=circle,draw,inner sep=1pt] (char) {#1};
	}\xspace
}

%% file: tex/acro.tex
\begin{acronym}
	
	\acro{ANN}{Artificial Neural Network}
	\acro{API}{Application Programming Interface}
		
	\acro{CIR}{Channel Impulse Response}
	\acro{CNN}{Convolutional Neural Network}
	\acro{CPS}{Cyber-physical System}
	\acro{CSI}{Channel State Information}
	\acro{CSM}{Collaborative Sensing Mechanism}
	\acro{CV}{Cross Validation}
	
	\acro{DNN}{Deep Neural Network}
	
	\acro{FPP}{First Path Power}
	
	\acro{HAR}{Human Activity Recognition}
	\acro{HT-LTF}{High Throughput LTF}
	
	\acro{IoT}{Internet of Things}
	\acro{ITS}{Intelligent Transportation System}
	
	\acro{LIMITS}{LIghtweight Machine Learning for IoT Systems}
	\acro{LOS}{Line of Sight}
	\acro{LTF}{Long-Training Field}
	\acro{LLTF}{Legacy Long Training Field}
	
	\acro{MCU}{Microcontroller Unit}
	\acro{ML}{Machine Learning}
	
	\acro{NIC}{Network Interface Controller}
	
	\acro{OFDM}{Orthogonal Frequency-Division Multiplexing}
	
	\acro{PCA}{Principal Component Analysis}
	\acro{PCB}{Printed Circuit  Board}
	\acro{PPA}{Peak Path Amplitude}
	
	\acro{RF}{Random Forest}
	\acro{RSSI}{Received Signal Strength Indicator}
	\acro{RTI}{Radio Tomographic Imaging}
	
	\acro{SC}{subcarrier}
	\acro{STBC-HT-LTF}{Space-Time Block Code HT-LTF}
	\acro{SVM}{Support Vector Machine}
	
	\acro{TMS}{Traffic Monitoring System}
	
	\acro{USB}{Universal Serial Bus}
	\acro{UWB}{Ultra-Wideband}	
	
	\acro{WDWS}{Wireless Detection and Warning System}
	\acro{WEKA}{Waikato Environment for Knowledge Analysis}
	\acro{WIM}{Weigh in Motion}
	\acro{WLAN}{Wireless Local Area Network}
	\acro{WSN}{Wireless Sensor Network}
	
\end{acronym}

%% file: tex/title.tex
\title{\paperTitle}

\ifacm
\else

	\title{\paperTitle}

	\ifdoubleblind
	\author{\IEEEauthorblockN{\textbf{Anonymous Authors}}
		\IEEEauthorblockA{Anonymous Institutions\\
			e-mail: Anonymous Emails}}
	\else
	\author{\IEEEauthorblockN{\textbf{\paperAuthors}}
		\IEEEauthorblockA{Communication Networks Institute,	TU Dortmund University, 44227 Dortmund, Germany\\
			e-mail: \paperEmails}}
	\fi
	
	\maketitle

\fi




%% file: tex/abstract.tex
\begin{abstract}
	
%
%
The advancing deployment of ubiquitous \ac{IoT}-powered vehicle detection and classification systems will successively turn the existing road infrastructure into a highly dynamical and interconnected \ac{CPS}. Though many different sensor systems have been proposed in recent years, these solutions can only meet a subset of requirements, including cost-efficiency, robustness, accuracy, and privacy preservation. 
%
%
This paper provides a modular system approach that exploits radio tomography in terms of attenuation patterns and highly accurate channel information for reliable and robust detection and classification of different road users. Hereto, we use \ac{WLAN} and \ac{UWB} transceiver modules providing either \ac{CSI} or \ac{CIR} data. Since the proposed system utilizes off-the-shelf and power-efficient embedded systems, it allows for a cost-efficient ad-hoc deployment in existing road infrastructures.
%
%
We have evaluated the proposed system's performance for cyclists and other motorized vehicles with an experimental live deployment. In this concern, the primary focus has been on the accurate detection of cyclists on a bicycle path. However, we also have conducted preliminary evaluation tests measuring different motorized vehicles using a similar system configuration as for the cyclists. In summary, the system achieves up to 100\% accuracy for detecting cyclists and more than 98\% classifying cyclists and cars. 
\end{abstract}

%% file: tex/arxiv.tex
\begin{tikzpicture}[remember picture, overlay]
\node[below=5mm of current page.north, text width=20cm,font=\sffamily\footnotesize,align=center] {Accepted for presentation in: 2021 Annual IEEE International Systems Conference (SysCon)\vspace{0.3cm}\\\pdfcomment[color=yellow,icon=Note]{
@InProceedings\{Haferkamp/etal/2021a,\\
  author    = \{Marcus Haferkamp and Benjamin Sliwa and Christian Wietfeld\},\\
  booktitle = \{2021 Annual IEEE International Systems Conference (SysCon)\},\\
  title     = \{A low cost modular radio tomography system for bicycle and vehicle detection and classification\},\\
  year      = \{2021\},\\
  address   = \{Vancouver, Canada\},\\
  month     = \{Apr\},\\
  publisher = \{IEEE\},\\
\}
}};
\node[above=5mm of current page.south, text width=15cm,font=\sffamily\footnotesize] {2021~IEEE. Personal use of this material is permitted. Permission from IEEE must be obtained for all other uses, including reprinting/republishing this material for advertising or promotional purposes, collecting new collected works for resale or redistribution to servers or lists, or reuse of any copyrighted component of this work in other works.};
\end{tikzpicture}

%% file: tex/introduction.tex
\section{Introduction}

%
%
Comprehensive and reliable \acp{ITS} are a crucial feature for emerging smart cities as the continuing increase in road traffic will noticeably exhaust the capacity of existing traffic systems~\cite{Sliwa/etal/2019b}. In many cases, constructional measures for expanding a traffic system's capacity are not an option, so traffic flow optimization is the only valuable solution resulting in data-driven \acp{ITS}. By continuously gathering specific information for different vehicle types, those systems enable more comprehensive traffic flow optimization than approaches providing only coarse indicators like traffic flow and traffic density.
%
%
Hence, those systems must meet several conditions at once, including a high detection and classification accuracy in real-time, even for challenging weather conditions. Moreover, they should provide energy-efficient, low-maintenance, and thus cost-efficient operation while being privacy-preserving. The compliance with those demands is highly relevant, particularly for mass deployments used in smart city applications. However, most of the existing solutions lack at least one of these criteria, disqualifying them for large-scale deployments.

%
%
\basicFig{b}{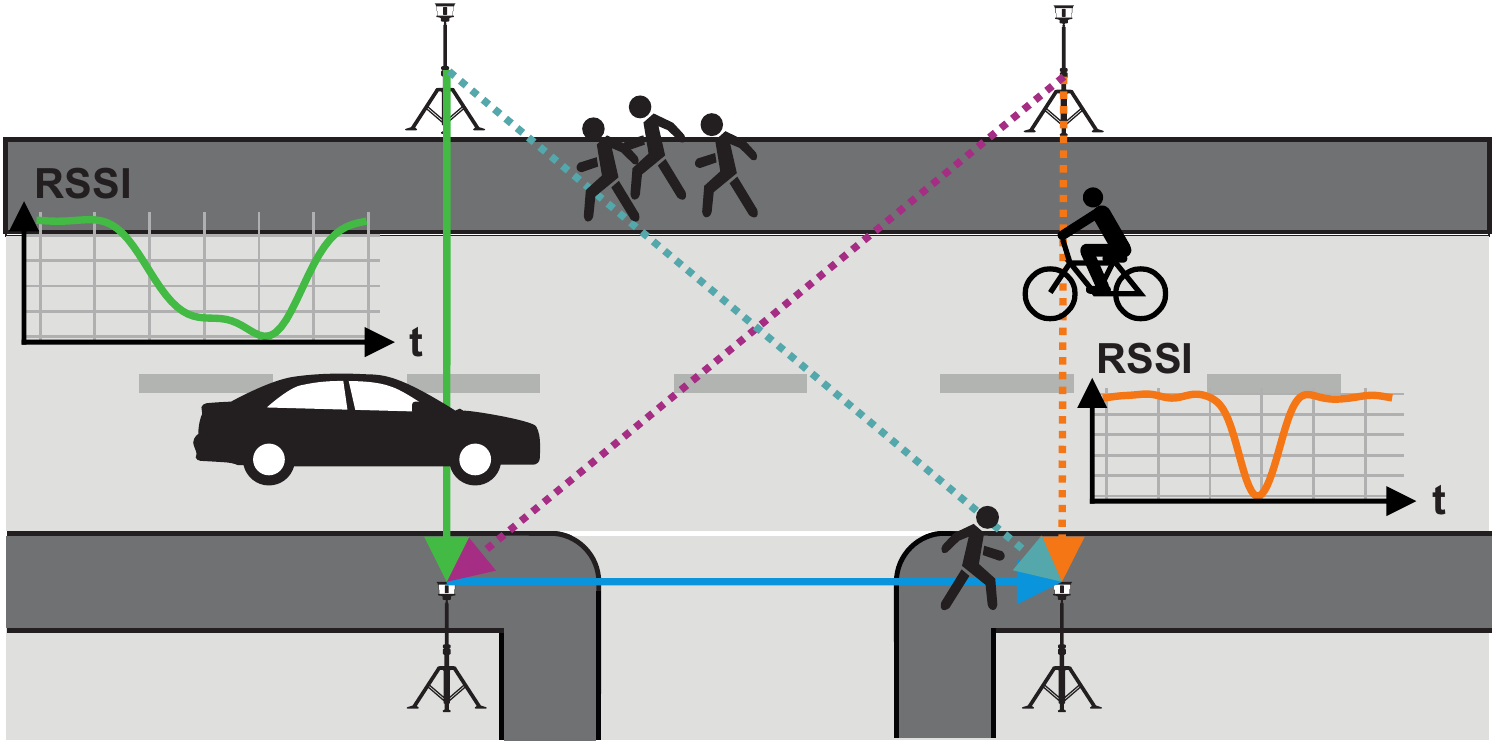}{Example application: Using the modular radio tomography system to detect and classify heterogeneous road users in an urban setting.}{fig:scenario}{0cm}{0cm}{1}
Hence, we present a modular and highly integrated \ac{WSN} installation for vehicle detection and classification that leverages both attenuation and high-dimensional channel information. The central assumption is that each vehicle induces type-specific radio channel patterns (\textit{fingerprints}), allowing for accurate vehicle detection and classification. Hereto, we use different state-of-the-art \ac{ML} models suitable for deployment to off-the-shelf \acp{MCU} for implementing a highly automated classification process. Thus, our system fulfills the previously mentioned requirements for smart city applications, i.\,e., high detection and classification accuracy, robustness against challenging weather conditions, cost-efficiency, and privacy-preservation. The initial \ac{WDWS} has exploited the attenuation of radio links induced by passing vehicles to detect wrong-way drivers on motorways~\cite{Haendeler2014}. Subsequently, this approach has been successively adopted for a fine-grained and \ac{ML}-based vehicle classification of multiple vehicle classes~\cite{Sliwa/etal/2020a}. 

This paper proposes a modular and highly integrated radio-based detection system, allowing for cost-efficient mass deployments in urban road infrastructure. As an example, Fig.~\ref{fig:scenario} illustrates the proposed system's use for automated detection and classification of cyclists and vehicles in an urban scenario.

The contribution of this paper is as follows: 
\begin{itemize}
	\item {Presentation of a low-cost, power-efficient, and \textbf{modular radio tomography system} for vehicle detection and classification exploiting highly accurate channel information}
	\item {Performance comparison of state-of-the-art \textbf{machine learning} methods---\ac{ANN}, \ac{RF}, \ac{SVM}---for two classification tasks}
	\item {In-depth suitability analysis of parameters extracted from \textbf{\ac{WLAN} \ac{CSI} and \ac{UWB} \ac{CIR} channel information}
	}
\end{itemize} 
%
%
After giving an overview of related work in Sec.~\ref{sec:related_work}, we provide the modular and radio-based classification system approach in Sec.~\ref{sec:approach}, the methodology in Sec.~\ref{sec:methods}, and present the performance analysis in Sec.~\ref{sec:performance_analysis}.

%
%

%
%
%

%% file: tex/relatedWork.tex
\section{Related Work}\label{sec:related_work}
In this section, we provide a brief overview of various sensor technologies used for vehicle detection and classification systems. Hereafter, we focus on related radio-based sensor systems and corresponding \ac{ML} models.

%
%
\basicFig{t}{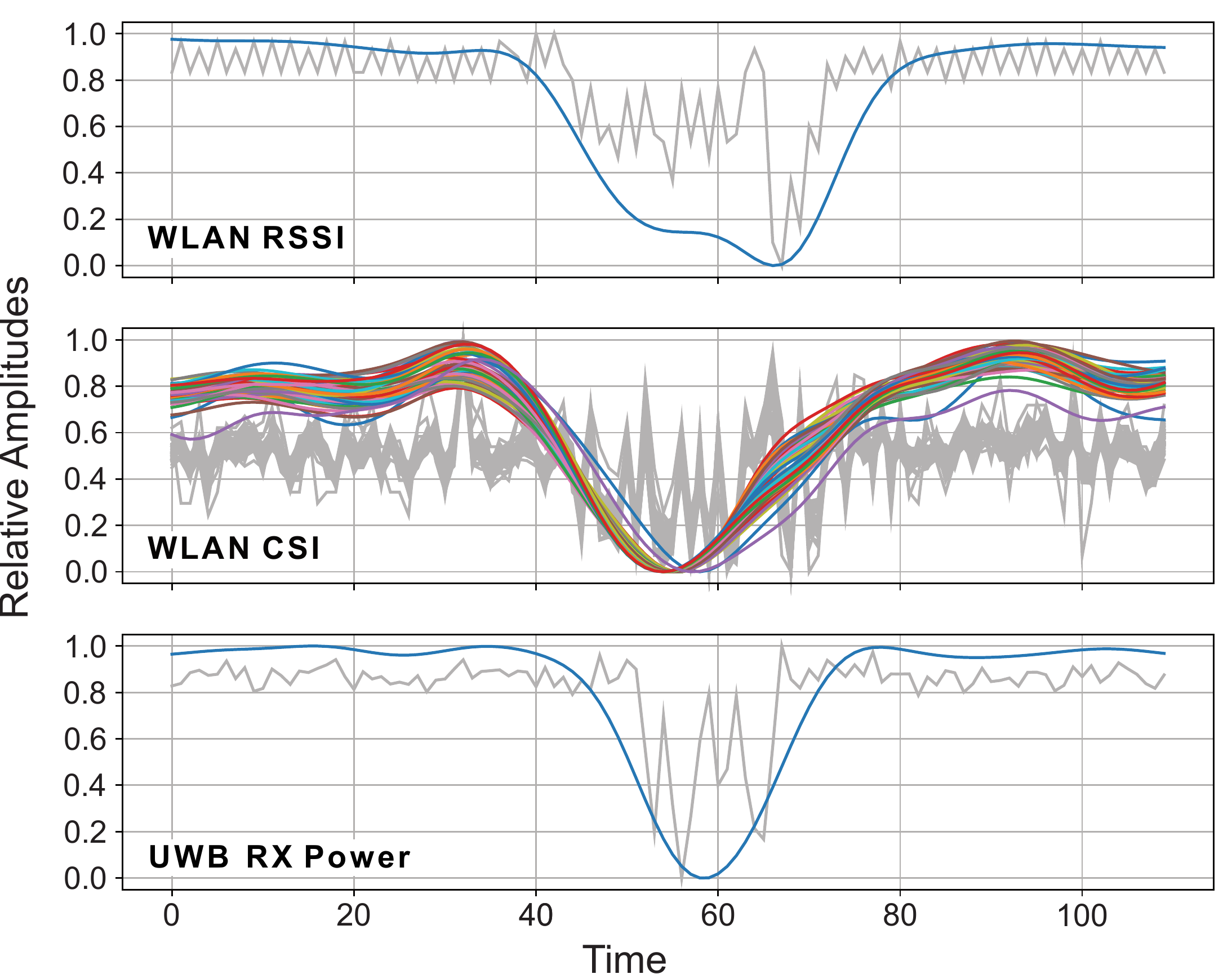}{Example raw and smoothed radio fingerprints of a cyclist for WLAN \ac{RSSI}, WLAN \ac{CSI}, and \ac{UWB} received signal power. For WLAN \ac{CSI}, each line indicates a subcarrier's relative amplitude. }{fig:attenuation_time}{0cm}{0cm}{1}

%
%
%

\subsection{Sensor Technologies for Detection and Classification}
Each vehicle detection and classification system can be classified either as \textbf{intrusive} or \textbf{non-intrusive}. While the former system type represents the \textit{original} system design and implies expensive roadwork for installation and maintenance (e.\,g., pavement cut), the latter is typically well-suited for large-scale deployments due to less extensive effort. 

Specifically, systems categorized as intrusive are: \ac{WIM}~\cite{Klein2006}, induction loops~\cite{Wu2014,Lamas2015}, fiber Bragg grating sensors~\cite{AlTarawneh2018}, vibration sensors~\cite{Ye2020}, and piezoelectric sensors~\cite{Rajab2016}. Contrary, there is a variety of non-intrusive sensor technologies used for detection and classification systems, which includes acoustic sensors~\cite{George2013, Daniel2016}, inertial sensors~\cite{Xu2018, Ma2014}, vision-based~\cite{Siddiqui2015, Liu2015} as well as radio-based systems. In the following, we discuss radio-based approaches in more detail.

\subsection{Radio-based Sensor Systems}
%
%
%
Radio-based approaches leverage \textit{radio tomography} and \ac{RTI}~\cite{Anderson/etal/2014a} for conducting detection and classification tasks. Such systems are \acp{WSN} ranging from simple one-link setups to collaborative multi-technology systems exploiting different radio technologies. The basic assumption of radio tomography is that objects of different shapes and materials lead to characteristic radio signal patterns. The resulting \textit{radio fingerprint} can be used for several kinds of object detection and tracking by taking snapshots over time (cf. Fig.~\ref{fig:attenuation_time}).

%
%
The \acf{RSSI} is a granular measure representing an estimate of the total received signal strength provided by most transceiver modules. For instance, the \ac{RSSI} is used in \acp{WSN} equipped with Bluetooth Low Energy beacons for vehicle detection and classification~\cite{Bernas2018}, achieving a detection and classification accuracy of up to 98\% and 97\% for three vehicle types, respectively. In~\cite{Sliwa/etal/2020a}, the authors propose an \ac{RSSI}-based multi-link vehicle classification system capable of conducting binary classifications with more than 99\% and more fine-grained seven-type classifications with more than 93\% accuracy assessing the \ac{RSSI} of each radio link. \\
%
%
In contrast to \ac{RSSI}, WLAN \ac{CSI} provides frequency-specific details regarding a radio channel. In general, \ac{OFDM}-based radio systems estimate \ac{CSI} for compensating a radio link's interferences to reconstruct the original symbols. In particular, the \ac{CSI} describes the estimated impact of the channel on both amplitude and phase of each subcarrier in the \ac{LTF} of a received packet. The total size of the \ac{CSI} depends on the number of transmit antennas, receive antennas, and subcarriers, whereas the latter varies between 64 and 512 subcarriers depending on the used channel bandwidth.

%
%
The great potential of \ac{CSI} becomes apparent when looking at various research activities. 
%
%
For instance, Adib et al. apply localization and tracking of moving objects behind a wall or closed doors. Furthermore, this approach also allows for detecting simple gestures performed behind a wall~\cite{Adib/Katabi/2013a}. Keenan et al. utilize this potential to distinguish three forms of human falling enabling privacy-preserving monitoring by healthcare applications. The proposed system achieves a balanced accuracy of 91\%, determining intended fall-like activities like sitting down and harmful ones such as walking-falls~\cite{Keenan2020}. Although \ac{UWB} is primarily used for indoor and outdoor localization, Sharma et al. compare the feasibility of \ac{WLAN} \ac{CSI} and \ac{UWB} for device-free \ac{HAR}~\cite{Sharma2019}. According to the presented results, \ac{UWB} outperforms \ac{WLAN} \ac{CSI} using an \ac{ML}-based classification for three different activities.

%
%
Concerning traffic monitoring, Won et al. leverage \ac{CSI} using two laptops equipped with \ac{WLAN} \acp{NIC}, detecting and classifying a total of five different vehicle types. The proposed classification system transforms the low-pass filtered and \ac{PCA}-treated \ac{CSI} data into image data, which serves as input for a \ac{CNN}, leading to average vehicle detection and classification accuracies of 99.4\% and 91.1\%, respectively~\cite{Won2019}. 

%
%
%
Instead of utilizing  only a single radio technology, Wang et al. propose a \ac{CSM}-based real-time vehicle detection and classification system combining power-efficient magnetic sensors and power-hungry cameras. While the low-cost magnetic sensors are running continuously for vehicle detection, the latter is usually in low-power mode and awake only for real-time vehicle classification and counting. This collaborative \ac{WSN} approach reaches a classification accuracy of at least 84\% for the vehicle types bicycle (98.84\%), car (95.71\%), and minibus (84.38\%)\cite{Wang/etal/2014a}.

%
%
Usually, \ac{CSI} is processed within the transceiver modules and, therefore, not directly accessible in most off-the-shelf devices. Hence, recent research has originated tools for extracting \ac{CSI} from specific \ac{WLAN} \acp{NIC}~\cite{Halperin2011, Xie2019}. However, using \texttt{Espressif ESP32} \acp{MCU} in our modular radio tomography system, we can directly access \ac{CSI} through the official firmware \ac{API}~\cite{Atif/etal/2020a}.

\begin{figure*}[t]  	
	\vspace{0cm}
	\centering		  
	\includegraphics[width=\textwidth]{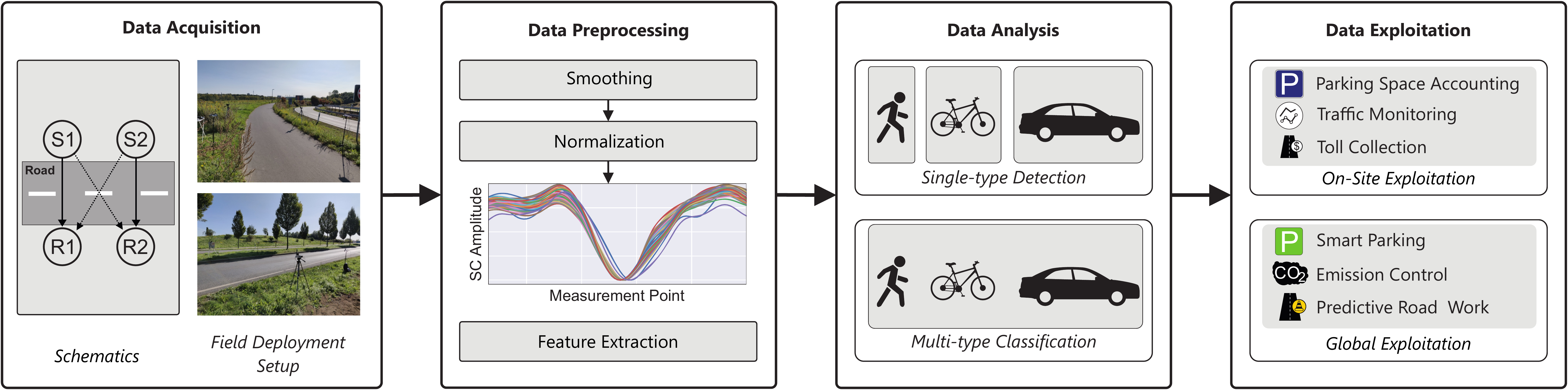}
	\caption{Overall system architecture model for a low-cost and modular radio tomography system for road user detection and classification. Radio fingerprints are gathered, preprocessed, evaluated using \ac{ML} algorithms, and exploited for different \ac{ITS} applications.}
	\label{fig:architecture}
	\vspace{0cm}	
\end{figure*}

\subsection{Machine Learning}
In recent years, the availability of numerous differently scaling \ac{ML} algorithms has promoted their use in many application areas, including the cognitive optimization of radio-based applications. 
%
For vehicle detection and classification, the focus is on \textit{supervised} learning models such as \ac{ANN}, \ac{RF}~\cite{Breiman/2001a}, and \ac{SVM}~\cite{Cortes/Vapnik/1995a}. In contrast, more modern and complex \ac{ML} approaches---such as \acp{DNN}---are used less frequently due to their demand for large datasets.
%
%
Moreover, \ac{ML} models perform differently, mainly depending on the number of considered vehicle classes, the system deployment's environment, and the used \ac{WSN}, differentiating in the number of links, sensor technologies, etc. 







%% file: tex/approach.tex
\section{Solution Approach}\label{sec:approach}
%

%
\basicFig{b}{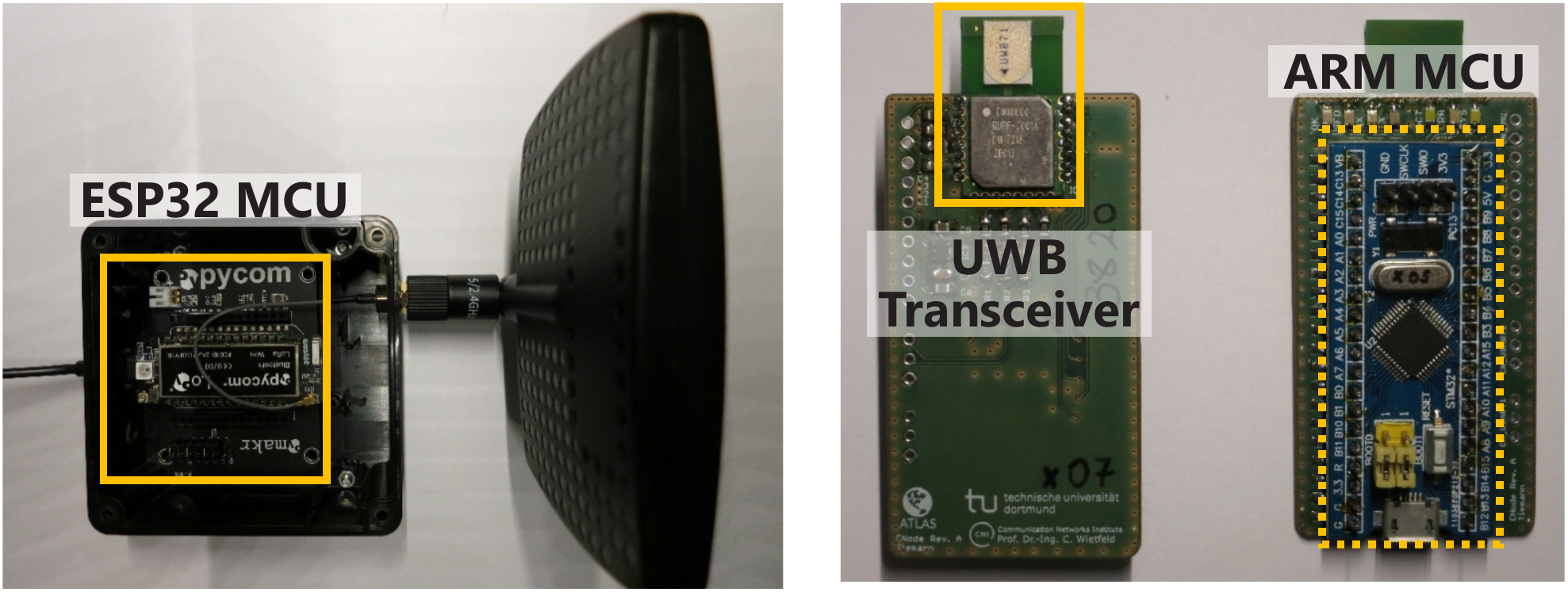}{\ac{WLAN} \ac{CSI} (left) and \ac{UWB} (right) transceiver modules evaluated in the low-cost and modular radio tomography system.}{fig:radio_modules}{0cm}{0cm}{1}
%

%
%
In this section, we explain the proposed solution approach and its components. For a better overview, Fig.~\ref{fig:architecture} illustrates the overall system architecture model containing four basic processing steps: data acquisition in the live system deployment, data preprocessing---including smoothing, normalization, and feature extraction---, \ac{ML}-based data analysis considering specific classification tasks, and data exploitation as required by various \ac{ITS} applications. 

%
%
\textbf{Data Acquisition:} Due to its data-driven nature, real-world traces of road users---e.\,g., bicycles and motorized vehicles---are gathered using a low-cost and modular radio-based \ac{WSN} setup. We evaluate two radio communication technologies: \ac{WLAN} \ac{CSI} and \ac{UWB} (cf. Fig~\ref{fig:radio_modules}). We utilize \texttt{Espressif ESP32} \acp{MCU} to access \ac{WLAN} \ac{CSI} and custom-made \acp{PCB}, combining a \texttt{Decawave DWM1000} \ac{UWB} transceiver module and an \texttt{ARM Cortex M3} \ac{MCU}~\cite{Tiemann2019}. Both \acp{MCU} provide the channel data via \ac{USB} interface for further processing.

%
%
%
\input{tex/tables/parameters}

%
%
%
%
\textbf{Data Preprocessing:} The raw \ac{WLAN} \ac{CSI} and \ac{UWB} \ac{CIR} data passes a three-step process cascade, including smoothing, normalization, and feature extraction. We conduct the data smoothing with a one-dimensional Gaussian filter evaluating different values for the Gaussian kernel's standard deviation $\sigma$. Hereafter, the smoothed data is normalized such that the values are bound to the range $[0,1]$ (\textit{min-max-scaling}). While we perform the smoothing to minimize the impact of scattered outliers---e.\,g., due to fading in the radio channel---the normalization enables high compatibility with the used \ac{ML} algorithms (\textit{feature scaling}). The last step is the extraction of multiple descriptive statistical features. In total, we have derived more than 20 attributes for the \ac{ML}-based classification.

%
%
\textbf{Data Analysis:} In the third process step, we feed the preprocessed data as input for two data analysis options. While one option targets the detection of only one specific vehicle type, the other one is required to detect and classify multiple vehicle types correctly. For instance, we performed the coarse-grained detection task along a cycle path counting cyclists. The latter application is more relevant for urban environments revealing heterogeneous road users, including pedestrians, cyclists, and several motorized vehicles. 

%
%
\textbf{Data Exploitation:} Finally, one could use the obtained data analysis results to provide multiple \ac{ITS}-related services either within a specific site (on-site exploitation) or on a large scale (global exploitation). Possible applications for on-site exploitation are parking space accounting, traffic monitoring, or toll collection. In contrast, analysis data acquired from multiple sensor deployments within a region can serve as input for smart parking, emission control, and predictive road work. 

%
%
%
%
%

%% file: tex/tables/parameters.tex
%
%
%
%

\begin{table}[b]
	\centering
	\caption{Main Parameters of the Proposed Modular Radio Tomography System Using \ac{WLAN} \ac{CSI} and \ac{UWB} for Bicycle Detection and Vehicle Classification.}
	\label{tab:parameters}
		\begin{tabular}{ccc}
			
			\toprule
			\multirow{2}{*}{\textbf{\normalsize{Parameter}}} &
			\multicolumn{2}{c}{\normalsize{\textbf{Radio Technology}}} \\
			& {\normalsize{\ac{WLAN} \ac{CSI}}} & {\normalsize{\ac{UWB}}} \\

			\midrule
			
			Transmission power& 20 dBm & 9.3 dBm \\
			Operating frequency & 2.4 GHz & 6.5 GHz \\
			Sampling frequency & 80 Hz & 40 Hz \\
			Antenna type & Directional & Omnidirectional \\
			Antenna gain & 5-7 dBi & --- \\
			Antenna height & 1m & 1m \\
			Number of radio links & 1 & 1 \\
			Distance TX $\leftrightarrow$ RX (cycle path) & 4m & 4m \\
			Distance TX $\leftrightarrow$ RX (road) & 7m & 7m \\
			
			\bottomrule
			
		\end{tabular}
\end{table}
%

%% file: tex/methodology.tex
\section{Methodology} \label{sec:methods}
This section provides details regarding the modular radio tomography system's parameters, the vehicle taxonomies assumed for the classification task, and in-depth information about the \ac{ML} models we have applied in the evaluation step. 

%
%
\subsection{Field Deployment Setup} \label{subsec:field_deployment}
Tab.~\ref{tab:parameters} summarizes the essential system parameters of the proposed radio-based detection and classification system. We have comparably installed \ac{WLAN} \ac{CSI} and \ac{UWB} transceiver modules in the field deployment setup. Nevertheless, some differences face the transmission power or the antenna characteristics induced either by the transceiver modules' design or the radio technology. Moreover, there is a variation concerning the distances between transmitter and receiving nodes for measuring cyclists and motorized vehicles. We have gathered radio fingerprints along a cycle path and a busy one-lane road, respectively. 
Since most captured fingerprints are related to cyclists (995 traces), this paper's primary focus is on detecting these—which can be interpreted as a binary classification of \textit{bicycle} and \textit{non-bicycle}. For this reason, we also have captured \textit{idle} traces, i.\,e., there is a \ac{LOS} between transmitter and receiver. Hereafter, we also evaluate the proposed system's applicability for a more fine-grained detection and classification task of three types: idle, cyclist, and car-like vehicles (cf. Fig.~\ref{fig:taxonomies}).

%
%
\basicFig{t}{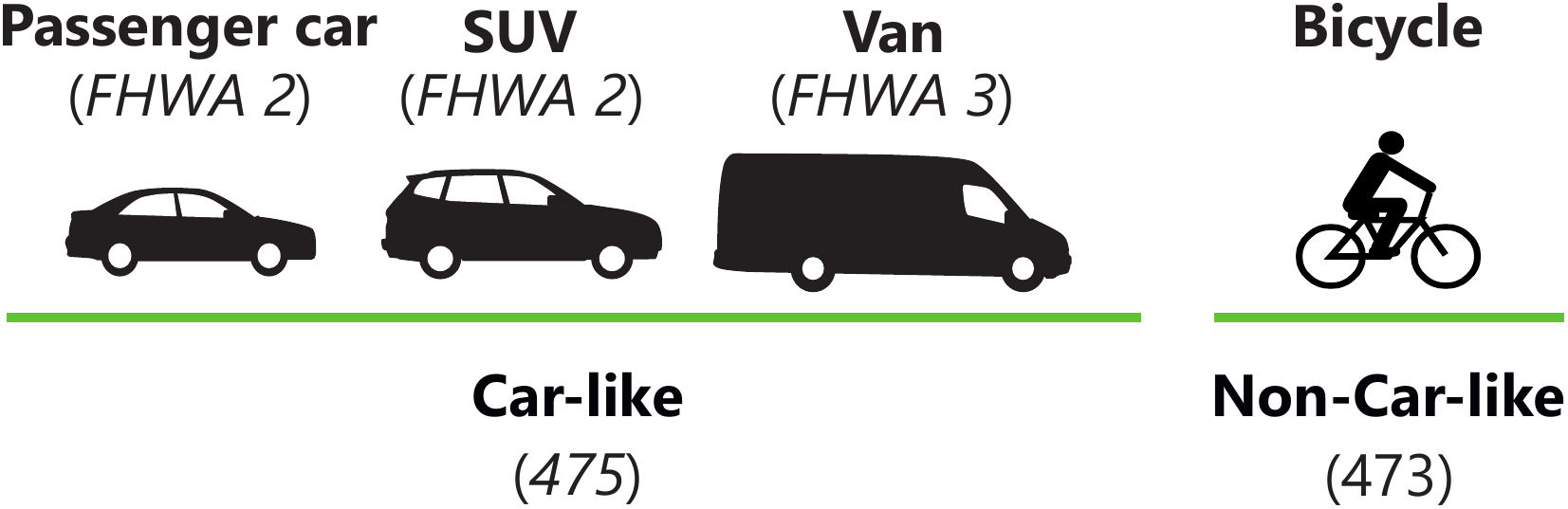}{Taxonomies: Vehicle classes and sample numbers used in the multi-type classification task. We considered balanced subsets for car-like, bicycle, and idle samples.}{fig:taxonomies}{-4pt}{0cm}{1.0}
%

%
%
\subsection{Machine Learning}
%
%
For the detection and classification, we utilize multiple models that have different implications for the achievable accuracy and resource efficiency. These considered models are chosen with respect to the findings of~\cite{Sliwa/etal/2020a}, which yielded that often less complex classification models achieve better accuracy results than cutting edge methods that would require a significantly higher amount of training data for achieving a comparable performance level due to the \emph{curse of dimensionality}.
%
%
%

%
%
\begin{itemize}
	%
	%
	\item \textbf{\acfp{ANN}}~\cite{LeCun/etal/2015a} aim to mimic core functions of the human nervous system and have received tremendous attention within various scientific communities in the context of \emph{deep learning}.	
	%
	%
	These models can be implemented as a sequence of matrix multiplications with element-wise node activations.
	%
	%
	The resulting memory size of \acp{ANN} is determined by their corresponding network architecture. Due to the usage of floating-point arithmetic, \acp{ANN} are less popular for being used on highly resource-constrained \ac{IoT} platforms such as ultra low power microcontrollers.

	%
	%
	\item \textbf{\acfp{RF}}~\cite{Breiman/2001a} are ensemble methods that base their decision making on the joint consideration of a number of random trees. Each tree is trained on a random subset of the features and a random subset of the training data. The layer-wise descent within the trees is based on binary decision making, whereas the value of a single feature is compared to a learned threshold.
	%
	%
	Due to condition-based decision making, \acp{RF} can be implemented in a highly resource-efficient manner as a sequence of \texttt{if/else} statements.
	%
	%
	Varying the number of trees and the maximum tree depth allows to control the memory usage of \acp{RF}.
	
	%
	%
	\item \textbf{\acfp{SVM}}~\cite{Cortes/Vapnik/1995a} learn a hyperplane for separating data points in a multidimensional space through minimization of a specific objective function. The hyperplanes are chosen for each feature that most members of one of two classes are on each of the hyperplane sides. We apply the \emph{one-vs-all} strategy for using \ac{SVM} for multi-class learning problems.

\end{itemize}
%
%
%

%
%
In order to assess the generalizability of the achieved classification results, we apply a $k=10$-fold cross-validation and investigate the variance of the model performance. Hereby, the overall data set $\mathcal{D}$ is divided into $k$ subsets $\{ \mathcal{D}_1, ..., \mathcal{D}_k \}$. In each iteration $i$, $\mathcal{D}_i$ is chosen as the training set $\mathcal{D}_{\text{train}}$ for the model, and the remaining subsets jointly compose the test set $\mathcal{D}_{\text{test}}$.

%
%
All data analysis tasks are carried out using the high-level \ac{LIMITS} framework~\cite{Sliwa/etal/2020c} for automating \ac{WEKA}~\cite{Hall/etal/2009a} evaluations. In addition, it allows exporting \texttt{C/C++} code of trained prediction models.

%% file: tex/results.tex
\input{tex/tables/bicycle_results}

\section{Performance Analysis} \label{sec:performance_analysis}

In this section, we discuss the results for bicycle detection and multi-type vehicle classification using the proposed modular radio tomography system. Essentially, we show the results for both the WLAN \ac{CSI} and the \ac{UWB} radio modules. 

%
%
\subsection{Bicycle Detection}
As mentioned in Sec.~\ref{subsec:field_deployment}, this paper's primary focus is on accurately detecting cyclists on a cycle path, i.\,e., differentiating \textit{bicycles} and \textit{non-bicycles} (idle). Nonetheless, we also provide results for a more fine-grained classification task in the following section. 
%
%
\basicFig{t}{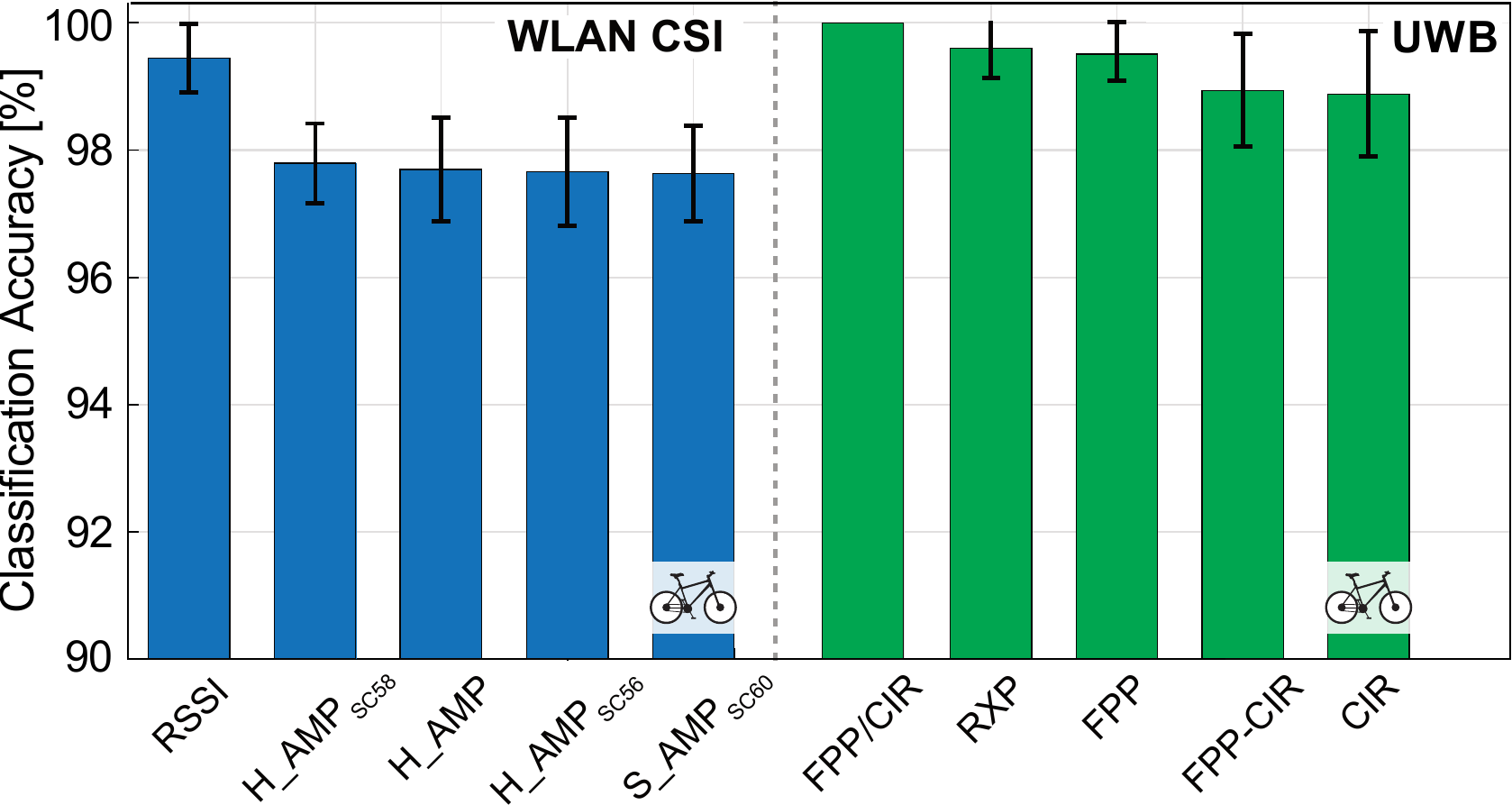}{Bicycle detection: Five most relevant channel parameters for \ac{WLAN} \ac{CSI} and \ac{UWB}, respectively. We evaluated each parameter separately using \ac{RF} and 10-fold \ac{CV}. \textit{\ac{CIR}}: \ac{CIR} power, \textit{\ac{FPP}}: First path signal power, \textit{H\_AMP}: Amplitudes of \ac{HT-LTF} subcarriers, \textit{\ac{RSSI}}: Received signal strength indicator, \textit{RXP}: Estimated received signal power, \textit{S\_AMP}: Amplitudes of \ac{STBC-HT-LTF} subcarriers, \textit{\ac{SC}}: Subcarrier.}{fig:channel_param_bike}{0cm}{0cm}{1}
Tab.~\ref{tab:bicycle_results} shows the classification results for the separately analyzed channel parameters acquired for WLAN \ac{CSI} and \ac{UWB} using the \ac{ML} models \ac{ANN}, \ac{RF}, and \ac{SVM}. Concerning WLAN \ac{CSI}, \textit{\ac{RSSI}} is the dominant channel parameter leading to the best classification results—for all scores. A possible explanation is that the WLAN transceiver module evaluates multiple channel parameters for calculating a single and significant indicator.
Similarly, one channel parameter is most relevant when using the \ac{UWB} transceiver modules: the quotient of the estimated \textit{\acf{FPP}} and the \textit{\acf{CIR} power}, where the latter is the sum of the magnitudes' squares from the estimated highest power portion of the channel. Using this extracted parameter \textit{\ac{FPP}/\ac{CIR}} and \ac{ANN}, we achieve a bicycle detection (binary classification) accuracy of 100\%. 

%
%
%
%
Fig.~\ref{fig:channel_param_bike} illustrates the five most relevant channel parameters of \ac{WLAN} \ac{CSI} and \ac{UWB} for bicycle detection using \ac{RF}. As previously discussed, the \textit{\ac{RSSI}} (\ac{WLAN} \ac{CSI}) and the quotient \textit{FPP/CIR} (\ac{UWB}) are the most significant channel parameters for correctly detecting cyclists. While the investigated \ac{UWB} parameters lead to small deviations regarding the classification accuracy, there is at least 2\% lower accuracy comparing \ac{RSSI} and the remaining WLAN \ac{CSI} parameters. 

%
%
\basicFig{b}{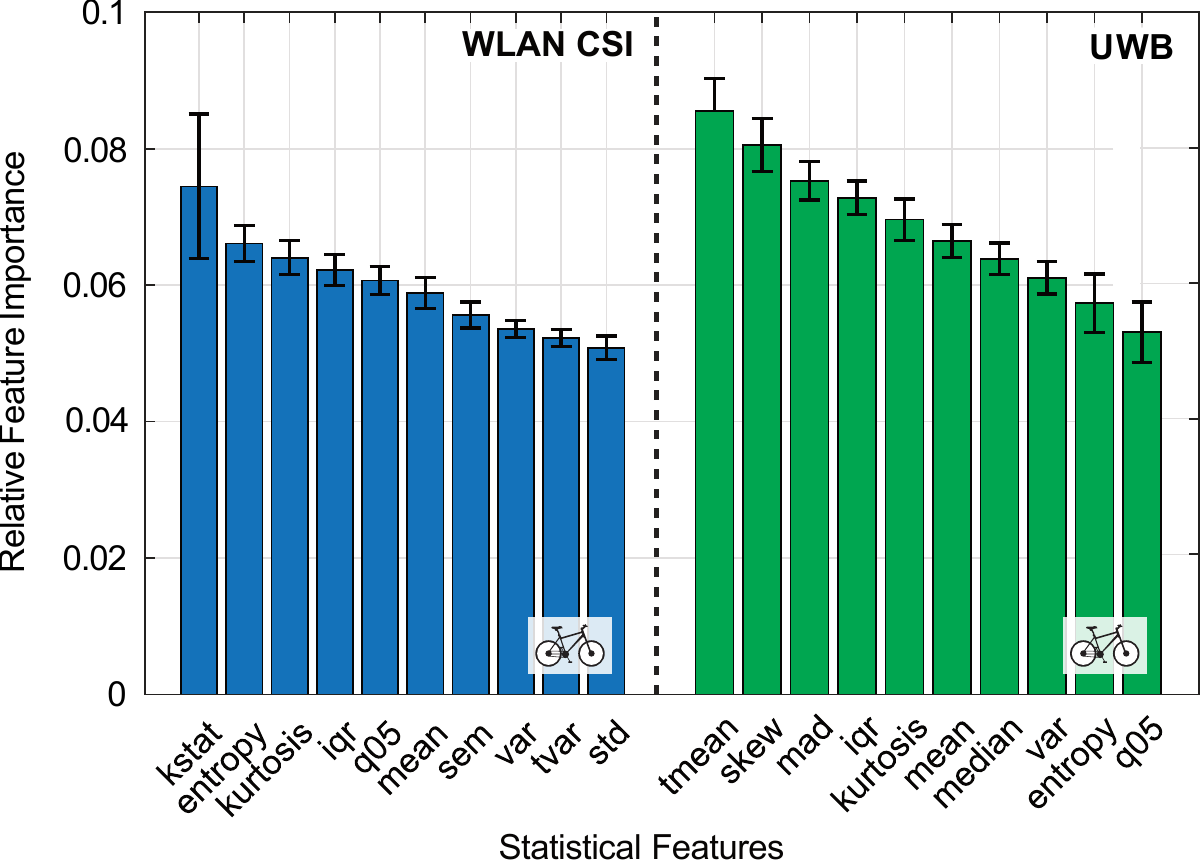}{Bicycle detection: Feature importance for the ten most relevant extracted statistical features for \ac{WLAN} \ac{CSI} and \ac{UWB} using \ac{RF}. \textit{iqr}: interquartile range, \textit{kstat}: k-static, \textit{mad}: median absolute deviation, \textit{q05}: $5^{th}$ quantile, \textit{q95}: $95^{th}$ quantile, \textit{sem}: standard error of mean, \textit{std}: standard deviation, \textit{tmean}: trimmed mean, \textit{tvar}: trimmed variance, \textit{var}: variance.}{fig:featureImportance_bike}{0cm}{0cm}{1}
%
%
%
%
\basicFig{t}{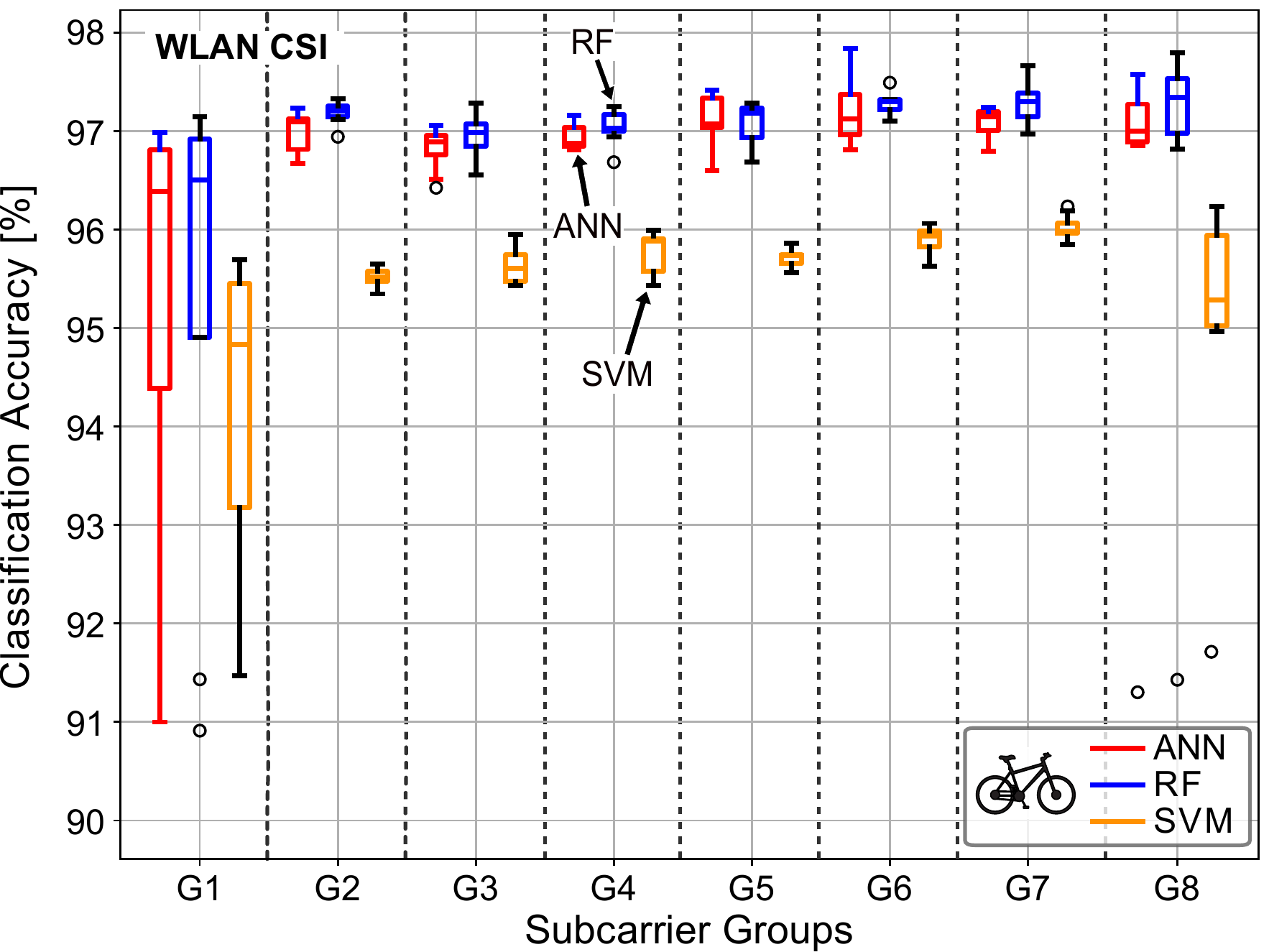}{Bicycle detection: Classification accuracy using different subcarrier amplitudes as input for \ac{ANN}, \ac{RF}, \ac{SVM}, respectively. For a better overview, we have grouped adjacent subcarriers.}{fig:sc_importance_bike}{0cm}{0cm}{1} 
Fig.~\ref{fig:featureImportance_bike} depicts the ten most significant extracted statistical features for \textit{\ac{RSSI}} and \textit{\ac{FPP}/\ac{CIR}}. For both systems, we can identify small differences in their relative feature importance distributions. Again for \ac{WLAN} \ac{CSI}, there is a single dominant feature (\textit{kstat}), whereas we cannot determine such a superior one regarding  \ac{UWB}.

%
%
%
%
Finally, Fig.~\ref{fig:sc_importance_bike} presents the significance of different \ac{WLAN} \ac{CSI} subcarrier amplitudes for the given binary classification task utilizing \ac{ANN}, \ac{RF}, and \ac{SVM}. For a better overview, we have split adjacent \acp{SC} into eight groups. We can state a frequency-specific relevance of these \acp{SC} regarding the classification accuracy. In particular, the \acp{SC} of \textit{G1} (\acp{SC} 1-8) are less suitable than those of the remaining groups. Furthermore, we can observe comparably high accuracies using \ac{ANN} and \ac{RF}, but consistently lower ones using \ac{SVM}. 
%
%
\subsection{Multi-Type Vehicle Classification}

This section provides an outlook on the modular radio system's applicability for multi-type vehicle classification. For a total of three evaluated categories---idle, bicycle (non-car-like), and car-like---Tab.~\ref{tab:bicycle_vehicle_results} lists the classification results for \ac{WLAN} \ac{CSI} and \ac{UWB} using \ac{ANN}, \ac{RF}, and \ac{SVM}, respectively. Contrary to the cyclist detection task, there are at least two predominant channel parameters for each system. 
\input{tex/tables/bicycle_vehicles_results}
Concerning \ac{WLAN} \ac{CSI}, the \ac{LLTF} subcarriers' amplitudes (\textit{L}) are most suitable using \ac{ANN}; instead, the \ac{STBC-HT-LTF} subcarriers' amplitudes (\textit{S}) are more crucial when applying \ac{RF}. There are two relevant parameters when using \ac{SVM}: the \ac{LLTF} subcarriers' amplitudes (\textit{L}) and the amplitudes of the $52^{nd}$ subcarrier in the \ac{HT-LTF} training field ($H_{SC52}$). 

Focusing on the classification results achieved for \ac{UWB}, there are also two major channel parameters: the amplitudes of all raw \ac{CIR} accumulator data (\textit{A}) and the amplitudes of accumulator sample 15 ($A_{15}$). When comparing the classification results for both systems, we can state a considerable performance gap for the benefit of \ac{WLAN} \ac{CSI}. We note that we have gathered traces of car-like vehicles on a busy one-lane road, implying a more substantial distance between sending and receiving nodes than for measuring cyclists, which may significantly affect the \ac{UWB} transceiver modules using omnidirectional antennas. 

\basicFig{b}{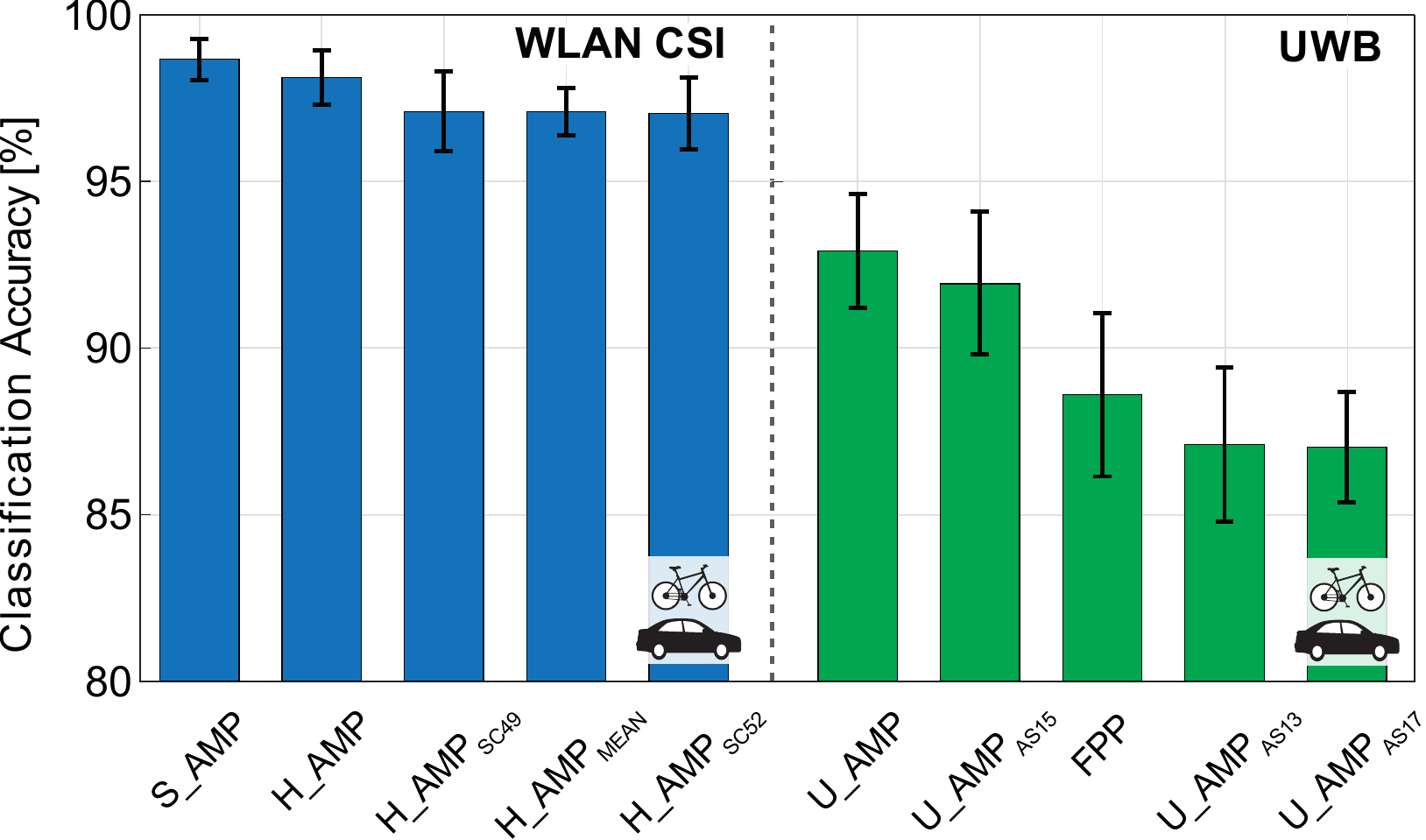}{Multi-type vehicle classification: Five most relevant channel parameters for \ac{WLAN} \ac{CSI} and \ac{UWB}, respectively. We evaluated each parameter separately using \ac{RF} and 10-fold \ac{CV}. \textit{AS}: Accumulator sample index, \textit{\ac{FPP}}: First path signal power, \textit{H\_AMP}: Amplitudes of \ac{HT-LTF} \acp{SC}, \textit{S\_AMP}: Amplitudes of \ac{STBC-HT-LTF} \acp{SC}, \textit{\ac{SC}}: Subcarrier, \textit{U\_AMP}: Amplitudes of \ac{CIR} accumulator samples}{fig:channel_param_bike_vehicle}{0cm}{0cm}{1}
Fig.~\ref{fig:channel_param_bike_vehicle} illustrates the relevance of different channel parameters gathered from \ac{WLAN} \ac{CSI} and \ac{UWB} regarding a three-type classification using \ac{RF}. Concerning the results, several \ac{WLAN} \ac{CSI} channel parameters lead to classification accuracies in the range of 97\% to 98\%. Contrary, the overall classification performance is notably worse, using any of the evaluated \ac{UWB} parameters. The most suitable \ac{UWB} parameter \textit{U\_AMP} results in about 4\% lower accuracy than the fifth most relevant \ac{WLAN} \ac{CSI} parameter $H\_AMP_{SC52}$. Furthermore, we achieve considerably different accuracy levels reaching from about 93\% down to 87\% using the five most relevant \ac{UWB} parameters. We assume that the divergent antenna types and sampling rates of the used \ac{WLAN} \ac{CSI} and \ac{UWB} transceiver modules (cf. Tab.~\ref{tab:parameters}) may cause this performance gap.

%% file: tex/tables/bicycle_results.tex
\begin{table}[b]
	\centering
	\caption{Bicycle Detection:  Results for \ac{WLAN} \ac{CSI} and \ac{UWB} using \ac{ANN}, \ac{RF}, and \ac{SVM} with a 10-fold \ac{CV}}
	\label{tab:bicycle_results}
	\begin{tabular}{clcccc}

		\multirow{2}{*}{{\small \textbf{Model}}} & \multirow{2}{*}{\textbf{\small{Score}}}
		&  \multicolumn{2}{c}{\small{\textbf{\ac{WLAN} \ac{CSI}}}} & \multicolumn{2}{c}{\small{\textbf{\ac{UWB}}}} \\
		& & {\scriptsize Value [\%]} & {\scriptsize Param.} & {\scriptsize Value [\%]} & {\scriptsize Param.} \\
		
		\midrule
		
		\multirow{4}{*}{\scriptsize{\ac{ANN}}} & \scriptsize{Accuracy} & \scriptsize{99.27$\pm$0.57} & {\scriptsize R (f2)} & \scriptsize{100$\pm$0} & {\scriptsize FC (f0)} \\
		& {\scriptsize Precision} & {\scriptsize 99.35$\pm$0.52} & {\scriptsize R (f2)}  & {\scriptsize 100$\pm$0} & {\scriptsize FC (f0)}  \\
		& {\scriptsize Recall} & {\scriptsize 99.24$\pm$0.61} & {\scriptsize R (f2)} & {\scriptsize 100$\pm$0} & {\scriptsize FC (f0)} \\ 
		& {\scriptsize F-Score} & {\scriptsize 99.30$\pm$0.56} & {\scriptsize R (f2)} & {\scriptsize 100$\pm$0} & {\scriptsize FC (f0)} \\
		\hline
		
		\multirow{4}{*}{{\scriptsize \ac{RF}}} & {\scriptsize Accuracy} & {\scriptsize 99.45$\pm$0.54} & {\scriptsize R (f0)} & {\scriptsize 99.83$\pm$0.26} & {\scriptsize FC (f1)} \\
		& {\scriptsize Precision} & {\scriptsize 99.48$\pm$0.52} & {\scriptsize R (f0)} & {\scriptsize 99.84$\pm$0.25} & {\scriptsize FC (f1)} \\
		& {\scriptsize Recall} & {\scriptsize 99.45$\pm$0.51} & {\scriptsize R (f0)} & {\scriptsize 99.8$\pm$0.26} & {\scriptsize FC (f1)} \\ 
		& {\scriptsize F-Score} & {\scriptsize 99.46$\pm$0.51} & {\scriptsize R (f0)} & {\scriptsize 99.83$\pm$0.26} & {\scriptsize FC (f1)} \\
		\hline
		
		\multirow{4}{*}{{\scriptsize \ac{SVM}}} & {\scriptsize Accuracy} & {\scriptsize 99.32$\pm$0.51} & {\scriptsize R (f2)} & {\scriptsize 99.83$\pm$0.26} & {\scriptsize FC (f0)} \\
		& {\scriptsize Precision} & {\scriptsize 99.38$\pm$0.47} & {\scriptsize R (f2)}  & {\scriptsize 99.84$\pm$0.24} & {\scriptsize FC (f0)} \\
		& {\scriptsize Recall} & {\scriptsize 99.30$\pm$0.53} & {\scriptsize R (f2)}  & {\scriptsize 99.82$\pm$0.27} & {\scriptsize FC (f0)} \\ 
		& {\scriptsize F-Score} & {\scriptsize 99.34$\pm$0.50} & {\scriptsize R (f2)} & {\scriptsize 99.83$\pm$0.26} & {\scriptsize FC (f0)} \\

		\bottomrule
		
	\end{tabular}
	\vspace{3pt}
	
	{\scriptsize \textit{f}: Filter size, \textit{FC}: Ratio of first path signal power and \ac{CIR} power, \textit{R}: RSSI } 
\end{table}

%% file: tex/tables/bicycle_vehicles_results.tex
\begin{table}[t]
	\centering
	\caption{Multi-type Vehicle Classification: Results for \ac{WLAN} \ac{CSI} and \ac{UWB} using \ac{ANN}, \ac{RF}, and \ac{SVM} with a 10-fold \ac{CV}}
	\label{tab:bicycle_vehicle_results}
	\begin{tabular}{clcccc}
		
		
		\multirow{2}{*}{\textbf{\small{Model}}} & \multirow{2}{*}{\textbf{\small{Score}}}
		&  \multicolumn{2}{c}{\small{\textbf{\ac{WLAN} \ac{CSI}}}} & \multicolumn{2}{c}{\small{\textbf{\ac{UWB}}}} \\
		& & {\scriptsize Value [\%]} & {\scriptsize Param.} & {\scriptsize Value [\%]} & {\scriptsize Param.} \\
		
		\midrule
		
		\multirow{4}{*}{\scriptsize{\ac{ANN}}} & \scriptsize{Accuracy} & \scriptsize{98.23$\pm$0.67} & {\scriptsize L (f4)} & \scriptsize{92.38$\pm$1.30} & {\scriptsize A (f2)} \\
		& {\scriptsize Precision} & {\scriptsize 98.52$\pm$0.49} & {\scriptsize L (f5)} & {\scriptsize 93.53$\pm$1.46} & {\scriptsize A (f2)} \\
		& {\scriptsize Recall} & {\scriptsize 98.31$\pm$0.63} &{\scriptsize L (f4)} & {\scriptsize 93.30$\pm$1.34} & {\scriptsize A (f2)} \\ 
		& {\scriptsize F-Score} & {\scriptsize 98.39$\pm$0.71} & {\scriptsize L (f3)} & {\scriptsize 93.41$\pm$1.38} & {\scriptsize A (f2)} \\
		\hline
		
		\multirow{4}{*}{\scriptsize \ac{RF}} & {\scriptsize Accuracy} & {\scriptsize 98.67$\pm$0.62} & {\scriptsize S (f0)} & {\scriptsize 92.96$\pm$1.67} & {\scriptsize A (f0)} \\
		& {\scriptsize Precision} & {\scriptsize 98.83$\pm$0.59} &{\scriptsize S (f0)} & {\scriptsize 93.74$\pm$1.74} & {\scriptsize A (f2)} \\
		& {\scriptsize Recall} & {\scriptsize 98.84$\pm$0.60} & {\scriptsize S (f1)} & {\scriptsize 93.28$\pm$1.79} & {\scriptsize A (f2)} \\ 
		& {\scriptsize F-Score} & {\scriptsize 98.8$\pm$0.61} & {\scriptsize S (f0)} & {\scriptsize 93.51$\pm$1.75} & {\scriptsize A (f2)} \\
		\hline
		
		\multirow{4}{*}{{\scriptsize \ac{SVM}}} & {\scriptsize Accuracy} & {\scriptsize 96.95$\pm$1.66} & {\scriptsize $\textnormal{H}_{\textnormal{SC52}}$ (f0)} & {\scriptsize 91.17$\pm$2.03} & {\scriptsize $\textnormal{A}_{\textnormal{15}}$ (f0)} \\
		& {\scriptsize Precision} & {\scriptsize 97.86$\pm$1.24} & {\scriptsize $\textnormal{H}_{\textnormal{SC52}}$ (f0)} & {\scriptsize 92.13$\pm$1.85} & {\scriptsize $\textnormal{A}_{\textnormal{15}}$ (f0)}  \\
		& {\scriptsize Recall} & {\scriptsize 97.46$\pm$0.43} & {\scriptsize L (f4)} & {\scriptsize 90.48$\pm$2.74} & {\scriptsize $\textnormal{A}_{\textnormal{15}}$ (f0)} \\ 
		& {\scriptsize F-Score} & {\scriptsize 97.39$\pm$1.44} &{\scriptsize $\textnormal{H}_{\textnormal{SC52}}$ (f0)} & {\scriptsize 91.29$\pm$2.25} & {\scriptsize  $\textnormal{A}_{\textnormal{15}}$ (f0)} \\

		\bottomrule
		
	\end{tabular}
	\vspace{2pt}
	
{\scriptsize 	{\textit{A}: Amplitudes of all \ac{CIR} accumulator samples, $A_{15}$: Amplitudes of \ac{CIR} accumulator sample 15, \textit{f}: Filter size, $H_{SC52}$: HT-LTF \ac{SC} 52 amplitudes, \textit{L}: LLTF \acp{SC} amplitudes, \textit{S}: STBC-HT-LTF \acp{SC} amplitudes
}}
	
\end{table}

%% file: tex/conclusion.tex
\section{Conclusion}

%
%
In this paper, we presented a novel bicycle detection and multi-type vehicle classification system that exploits highly accurate channel parameters provided by \ac{WLAN} \ac{CSI} and \ac{UWB}. Compared to existing traffic detection and classification systems, the proposed modular radio tomography system is privacy-preserving, robust against challenging weather conditions, and cost-efficient. Using real-world data from extensive field measurements, we have analyzed its applicability for two classification tasks with different state-of-the-art machine learning models. 
%
%
Regarding the detection of cyclists, which we conducted as a binary classification task, an accuracy of more than 99\% can be achieved for both radio technologies \ac{WLAN} \ac{CSI} and \ac{UWB}, using \ac{ANN}, \ac{RF}, and \ac{SVM}, respectively. Furthermore, we have evaluated the proposed system’s performance for a multi-type classification gaining more than 98\% accuracy. 

%
%
In future work, we will improve the system’s accuracy by correlating multiple radio links and extracting different radio channel parameters. Moreover, we will obtain additional samples of various vehicles involving challenging urban settings---e.\,g., in a downtown area with groupings of vehicles---and different weather conditions to strengthen the overall system performance. In the long term, the full detection and classification process, including the process steps discussed in this paper, should run self-sufficiently on the utilized \acp{MCU}. 

%% file: tex/acknowledgment.tex
\ifdoubleblind

\else

	\section*{Acknowledgment}
	
	\footnotesize
	This work has been supported by the PuLS project (03EMF0203B) funded by the German Federal Ministry of Transport and Digital Infrastructure (BMVI) and the German Research Foundation (DFG) within the Collaborative Research Center SFB 876 ``Providing Information by Resource-Constrained Analysis'', projects A4 and B4. We would like to thank Tugay Onat for his helpful support conducting the field measurements. 

\fi